\documentstyle[mnextra]{mn}

\seceqnum	

\input{epsf}
\newdimen\inmarginw \inmarginw=1.5 truecm 
\def\strutdepth{\dp\strutbox}
\def\inmargin#1{\strut\vadjust{\kern-\strutdepth{\vtop to
\strutdepth{\baselineskip%
\strutdepth\vss\llap{\hbox to \inmarginw{#1\hss}\space}\null}}}}

\def\eg{e.g.\ }
\def\ie{i.e.\ }
\def\lta{\mathrel{\hbox{\rlap{\hbox{\lower3pt\hbox{$\sim$}}}\hbox{$<$}}}}
\def\gta{\mathrel{\hbox{\rlap{\hbox{\lower3pt\hbox{$\sim$}}}\hbox{$>$}}}}
\def\lsim{\mathrel{\hbox{\rlap{\hbox{\lower3pt\hbox{$\sim$}}}\raise1pt\hbox{$<$}}}}
\def\gsim{\mathrel{\hbox{\rlap{\hbox{\lower3pt\hbox{$\sim$}}}\raise1pt\hbox{$>$}}}}

\def\ds{{\bf d}^{\rm S}}
\def\dl{{\bf d}^{\rm L}}
\def\Ds{\Delta^{\rm S}}
\def\Dl{\Delta^{\rm L}}
\def\delxs{\delta^{\rm S}({\bf x})}
\def\delks{{\delta}_{\bf k}^{\rm S}}
\def\delkl{{\delta}_{\bf k}^{\rm L}}
\def\q{{\bf q}}
\def\e{{\rm e}}
\def\k{{\bf k}}
\def\x{{\bf x}}
\def\v{{\bf v}}
\def\xp{{\bf x}^\prime}
\def\vp{{\bf v}^\prime}
\def\xpp{{\bf x}^{\prime\prime}}
\def\vpp{{\bf v}^{\prime\prime}}
\def\NS{N_{\rm S}}
\def\NL{N_{\rm L}}
\def\Mpc{ h^{-1}{\rm Mpc}}
\def\kpc{ h^{-1}{\rm kpc}}
\def\kms{ {\rm km \, s}^{-1} }
\def\div{{\bf \nabla} .}
\def\app{a^{\prime\prime}}
\def\TB{Tormen \& Bertschinger\ }
\def\eg{{\it e.g.\ }}
\def\ie{{\it i.e.\ }}
\def\etal{{\it et al.\ }}

\begin{document}

\title[Adding Long-Wavelength Power]{Adding Long-Wavelength Power to N-body Simulations}

\author[S. Cole ]{Shaun Cole\\
Department of Physics, University of Durham, Science
Laboratories, South Rd, Durham DH1 3LE.\\
Shaun.Cole@durham.ac.uk}

\maketitle

\begin{abstract}
\TB have presented an algorithm which allows the dynamic range of N-body
simulations to be extended by adding long-wavelength power to an
evolved N-body simulation.  This procedure is of considerable
interest as it will enable mock galaxy catalogues to be constructed
with volumes as large as those of the next generation of galaxy
redshift surveys. Their algorithm, however, neglects the coupling
between long-wavelength linear modes and short-wavelength non-linear
modes.  The growth of structure on small scales is coupled to the
amplitude of long-wavelength density perturbations via their effect on
the local value of the density parameter $\Omega_0$.  The effect of
neglecting this coupling is quantified using a set of specially
tailored N-body simulations.  It is shown that the large-scale
clustering of objects defined in the evolved density field such as
galaxy clusters is strongly underestimated by their algorithm.  An
adaptation to their algorithm is proposed that, at the expense of
additional complexity, remedies the shortcomings of the original
one.  Methods of constructing biased mock galaxy catalogues
which utilise the basic algorithm of Tormen \& Bertschinger, 
but avoid the pitfalls are discussed.
\end{abstract}


\section{Introduction}

\TB \shortcite{tb} have presented a procedure to extend
the dynamic range of large cosmological N-body simulations. Their
algorithm allows a periodic N-body simulation in a box of side $S
\gsim 100 \Mpc$ to be replicated and distorted to produce a particle
distribution in a larger box of side $L \gsim 800 \Mpc$, in which the
required linear power spectrum is accurately represented in the mass
distribution up to wavelengths $\lambda=L$.  This procedure can
therefore be used to take N-body simulations which have the mass
resolution to resolve small galaxy groups and from them construct mock
galaxy catalogues equal in volume to the next generation of redshift
surveys, \eg the Anglo-Australian 2df galaxy redshift survey 
(Maddox \etal in preparation; 
http: //msowww.anu.edu.au/~colless/2dF/) and the SDSS \cite{sdss}. 
Their method, however, neglects coupling between long-wavelength
linear modes and short-wavelength non-linear modes.
This coupling has only a small effect on the large scale clustering
properties of the mass distribution, but a much larger effect on
the large scale clustering of non-linear objects, 
which are highly biased tracers of the mass distribution
such as galaxy clusters or perhaps galaxies. 
In this paper we quantify the effect of neglecting 
this particular form of mode coupling and suggest how the algorithm
could be improved or its shortcomings circumvented.

  In Section~\ref{sec:LWP} we present the details of the algorithm
used to add long-wavelength power to an N-body simulations. A version
of the basic algorithm developed by \TB \shortcite{tb} is detailed in
Section~\ref{sec:space}. The effect of long-wavelength linear modes on the
evolution of short-wavelength non-linear modes is discussed in
Section~\ref{sec:time}, where a modification of the basic algorithm is
proposed that at the expense of additional complexity incorporates, to
first order, the effect of this form of mode coupling.  In
Section~\ref{sec:nbody} we use a set of specially tailored N-body
simulations to test the basic and extended algorithms and illustrate 
the effect of ignoring mode coupling.  In Section~\ref{sec:mgc} we address 
the specific problem of constructing large mock galaxy catalogues in which
galaxies are biased tracers of the mass distribution. We show how the
shortcomings of the basic algorithm can be largely circumvented by
restriction to a particular class of biasing scheme or by making a 
very minor modification to the basic  \TB algorithm. We conclude in
Section~\ref{sec:disc}.


\section{Adding Long-Wavelength Power}\label{sec:LWP}

The procedure described by Tormen \& Bertschinger \shortcite{tb} 
and named MAP (Mode Adding Procedure) consists
of the following steps.  First the power in long-wavelength modes is
removed from the original N-body particle positions and
velocities. This new particle distribution is then replicated to
define a particle distribution over a much larger volume than that of
the original simulation. A new density field is then generated with
the required power spectrum on a finer grid in $\k$-space, but populating
the same physical region of $\k$-space as the original modes that were removed.
The displacements computed from this grid are then used to perturb the
replicated particle positions and velocities. 

An implementation of these steps is described in
Section~\ref{sec:space}, while an additional procedure to deal with
the mode coupling not considered by Tormen \& Bertschinger
\shortcite{tb} is described in Section~\ref{sec:time}.


\subsection{Basic Algorithm: Spatial Perturbations}\label{sec:space}

The starting point of the MAP  are the final positions, $\x$, 
and velocities, $\v$, of a large periodic N-body simulation.
The size of the box, $S$, must be greater $\sim 100 \Mpc$ so
that the evolution of the longest wavelength modes present in the box 
is still described well by linear theory. One also requires the 
corresponding linear density field, $\delxs$, but since only the 
long-wavelength  contribution to this field is used this can be 
computed accurately from the particle positions.

The first step is to compute the long-wavelength contribution, $\Ds(\x)$,
to the density field $\delxs$. This is achieved by Fourier transforming
the density field and then constructing $\Ds(\x)$ using only 
the long-wavelength Fourier modes,
\begin{equation}
\Ds(\x) = \sum_{l,m,n} \delks  \ \e^{i \k.\x}
\end{equation}
where $ \k = (2\pi/S) (l, m, n) $ and $-\NS\le l,m,n \le \NS$.
The extent of the region of $\k$-space defining $\Ds(\x)$ (chosen
here for simplicity to be a cube) should be limited such that
$\langle\vert\Ds(\x)\vert^2\rangle \ll 1$. 
In practice this stage of the algorithm works
accurately provided $\langle\vert\Ds(\x)\vert^2\rangle^{1/2} \lsim 0.2$.
The next step is to compute the displacements produced by these modes 
using the Zel'dovich  \shortcite{zel} approximation,
\begin{equation}
\ds(\x) = \sum_{l,m,n} \frac{ \delks \,\k} {k^2} \ \e^{i \k.\x} .
\end{equation}
These displacements are readily computed on a grid using an FFT.
The removal of long-wavelength power from the  N-body simulation
is then achieved by subtracting these displacements from the final
N-body particle positions 
\begin{equation}
\xp = \x - \ds(\x)  
\label{eqn:xremove}
\end{equation}
and velocities
\begin{equation}
\vp = \v - f(\Omega) \, \ds(\x) .
\label{eqn:vremove}
\end{equation}
Here $f(\Omega$) is the
logarithmic derivative of the linear growth factor $D(a)$ with respect
to the expansion factor $a$ and if we use the convention that
positions are measured in units of $\Mpc$ then the velocities are
in units of $100\, \kms$. Note that the value of the field $\ds(\x)$ needs to be 
smoothly interpolated from the grid on which it is defined to each particle 
position.

Normally, when using the Zel'dovich approximation for a particle at
position $\x$, one computes the displacement $\ds(\q)$ evaluated at
the Lagrangian or starting position, $\q$, of that particle. Here
instead we have deliberately computed the displacement $\ds(\x)$ at the
final position $\x$.  For this reason the above procedure to remove
the long-wavelength power is not exact. However, provided the
displacements $\ds(\x)$ are small compared to the wavelength of the
modes comprising $\Ds(\x)$ the difference between $\ds(\q)$ and
$\ds(\x)$ is small and approximation works quite accurately.  This
constraint is identical to requiring
$\langle\vert\Ds(\x)\vert^2\rangle \ll 1$, as $\Ds(\x) = \div \ds(\x)$,
and so is well satisfied in all cases of interest.  Tormen \&
Bertschinger \shortcite{tb} describe a more elaborate and accurate
method of removing the long-wavelength power.  The reason for
deliberately using $\ds(\x)$ in equations (\ref{eqn:xremove}) and
(\ref{eqn:vremove}) is that we do not wish to disturb the small scale
structure of the N-body simulation. For example, we wish the
effect of the displacement field on a dense virialized cluster to be
to move the whole cluster but not disrupt its internal structure. Thus the
variation of displacement $\ds(\x)$ across the cluster should be
small. With the above formulation this occurs naturally as the
cluster particles span only a small range in $\x$,
but the shear would be considerably larger if $\ds(\q)$
were used since the cluster particles span a much larger volume
in the initial Lagrangian space.

The simulation, now with no power in long-wavelength modes, is 
replicated to define particle positions and velocities in a much
larger box of side $L \Mpc$. It is convenient to make $L$ an
odd multiple of $S$ so that in $\k$-space the two grids have cells
whose boundaries are aligned. A new Gaussian random field
is generated from the required power spectrum
\begin{equation}
\Dl(\x) = \sum_{l,m,n} \delkl \ \e^{i \k.\x}
\end{equation}
where $ \k = (2\pi/L) (l, m, n) $
and $-\NL\le l,m,n \le \NL$ with $2\NL+1 = (2\NS+1) (L/S)$.
Note that the sampling of $\k$-space is now on a much finer grid than
for $\Ds(\x)$.
The corresponding Zel'dovich approximation displacements, defined 
over the whole box of size $L \Mpc$, are given by
\begin{equation}
\dl(\x) = \sum_{l,m,n} \frac{ \delkl \, \k} {k^2} \ \e^{i \k.\x} .
\end{equation}
These are then simply added to the previous positions and velocities,
\begin{eqnarray}
\xpp &=& \xp + \dl(\xp)  \\ \label{eqn:xadd}
\vpp &=& \vp + f(\Omega) \, \dl(\xp)
\end{eqnarray}
to produce a particle distribution with the required improved sampling
of long-wavelength modes. Note that for this second stage of the algorithm
to work accurately one requires $\langle\vert\Dl(\x)\vert^2\rangle \ll 1$.
This is a more stringent condition than 
$\langle\vert\Ds(\x)\vert^2\rangle \ll 1$ as $\Dl(\x)$ contains contributions
from longer wavelength modes than are present in $\Ds(\x)$. In practice the
full algorithm works accurately if the region of $k$-space from which the modes
are removed and then re-sampled 
is such that $\langle\vert\Dl(\x)\vert^2\rangle^{1/2} \lsim 0.2$.


\subsection{Temporal Perturbations}\label{sec:time}

The addition of long-wavelength fluctuations in the initial
conditions of an N-body simulation has two effects. First
the long-wavelength modes produce a wave of expansion/compression 
which causes a large scale modulation in the initial particle
displacements and velocities. Second this same wave of expansion/compression 
varies the local mean density around each particle, thus perturbing
the local value of the density parameter $\Omega$.
While these long-wavelength perturbations are still in the linear regime
the first of these two effects has very little influence on the evolution
of small scale structure. For this reason its effect can
be accurately reproduced by the MAP described in Section~\ref{sec:space}, 
which applies these perturbations after the simulation has been evolved.
However the second effect of modulating the local value of the density 
parameter does influence the evolution of small scale structure,
because in a high $\Omega_0$ universe the linear growth rate of density
fluctuations, $D(a)$, grows faster than in a low $\Omega_0$ universe.
Thus where $\Omega_0$ is locally enhanced (diminished) by the
long-wavelength perturbations, clustering will take place faster (slower) 
than average. This effect is not reproduce by the MAP. Below we describe
a procedure whereby different regions of a simulation are allowed to
evolve for different times so as to mimic the effect that 
the long-wavelength density perturbation would have had on the
linear growth factor.

To first order the addition of a long-wavelength density perturbation, 
$\Delta(\x)$, perturbs the local value of the density parameter from 
$\Omega_0$ to $\Omega_0(1+\Delta(\x))$.  Since the
growth factor is also a function of time, or expansion factor, this
perturbation can also be produced by a temporal perturbation. 
Namely, in a region where one is adding a long-wavelength density perturbation,
$\Delta(\x)$, the simulation should be allowed to evolve longer to the point
where $D(\app,\Omega_0,\Lambda_0)=D(a,\Omega_0(1+\Delta),\Lambda_0)$.
Here $\Lambda_0$ is the cosmological constant.
Expanding both sides of this equation in a first order Taylor series,
\begin{eqnarray}
	&&D(a,\Omega_0,\Lambda_0) \left[1 + \frac{\app-a}{a}
\left(\frac{\partial \ln D} {\partial \ln a} \right)_{\Omega_0,\Lambda_0}
\right] \nonumber \\
&&= 
	D(a,\Omega_0,\Lambda_0) \left[1 + \Delta(\x)
\left(\frac{\partial \ln D} {\partial \ln \Omega_0} \right)_{a,\Lambda_0}
\right] ,
\end{eqnarray}
which simplifies to
\begin{equation}
\app \approx
      a \, \left[ 1 + \frac{\Delta(\x) }{f(\Omega)}
\left(\frac{\partial \ln D} {\partial \ln 
\Omega} \right)_{a,\Lambda_0} \right] 
\label{eqn:time}
\end{equation}
where $\Delta(\x)=\Dl(\x)-\Ds(\x)$.
For $\Lambda_0=0$ this reduces to
\begin{equation}
\app \approx a \, \left[ 1 +
\frac{\Delta(\x) (1-\Omega_0^{0.6})}  {(1-\Omega_0)\Omega_0^{0.6}} \right]
\end{equation}
where the usual approximation $f(\Omega)  \approx \Omega^{0.6}$ 
(\eg Peebles 1993) has been made. While for $\Omega_0=1$ this reduces to
simply
\begin{equation}
\app \approx a \, \left[ 1 + 0.6 \, \Delta(\x)  \right] .
\end{equation}

This procedure can be combined with the MAP described in
Section~\ref{sec:space} to define 
new particle positions and velocities 
\begin{eqnarray}
\xpp(a) = \x(\app) &-& \ds(\x(\app))  + \dl(\xp(\app)) 
\\
\label{eqn:xplusadd}
\vpp(a) = \v(\app) &-& f(\Omega_0) \ds(\x(\app))  \nonumber \\
&+& f(\Omega_0) \dl(\xp(\app)),
\label{eqn:vplusadd}
\end{eqnarray}
where $\app$ is given by (\ref{eqn:time}) and $\xp$ by (\ref{eqn:xremove}).
The practical problem with this procedure is that the N-body particle
distributions are now required not just at one single output time
but over a quite extended range. An efficient implementation 
requires merging the code to solve equation (\ref{eqn:time})
with the basic N-body evolution code so that individual particle positions
and velocities are only output at the required times. In order to create
catalogues spanning very large volumes it would be necessary also to merge
the code to carry out the transformations (\ref{eqn:xplusadd}) and 
(\ref{eqn:vplusadd}) and apply the galaxy catalogue selection function. If 
this were not done the replicated output of the N-body code would be 
unmanageably large.


\begin{figure*}
\centering
\centerline{\epsfxsize = 16.0 cm \epsfbox[0 250 574 774]{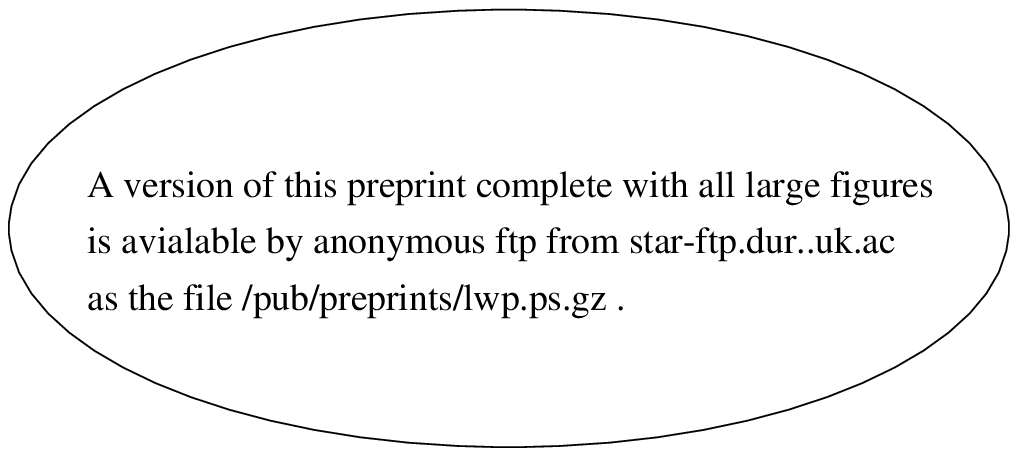}}
\caption{The mass distribution in a $5 \Mpc$ thick slice through the 
Prior simulation.}
\label{fig:prior}
\end{figure*}

\begin{figure*}
\centering
\centerline{\epsfxsize = 16.0 cm \epsfbox[0 250 574 774]{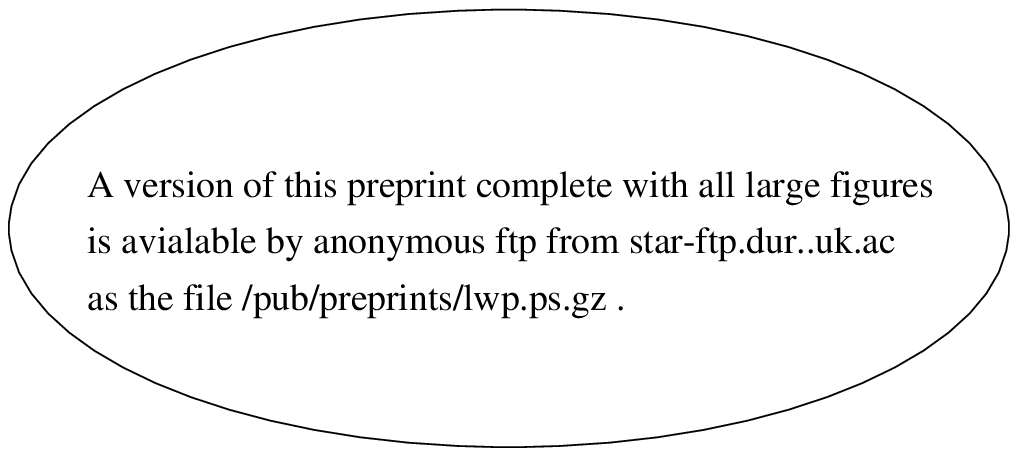}}
\caption{The mass distribution in the same slice as Fig.~\ref{fig:prior}, 
but for the Post simulation where long-wavelength power has been added to the 
Periodic simulation using the \TB (1996) procedure.  }
\label{fig:post}
\end{figure*}

\begin{figure*}
\centering
\centerline{\epsfxsize = 16.0 cm \epsfbox[0 0 574 774]{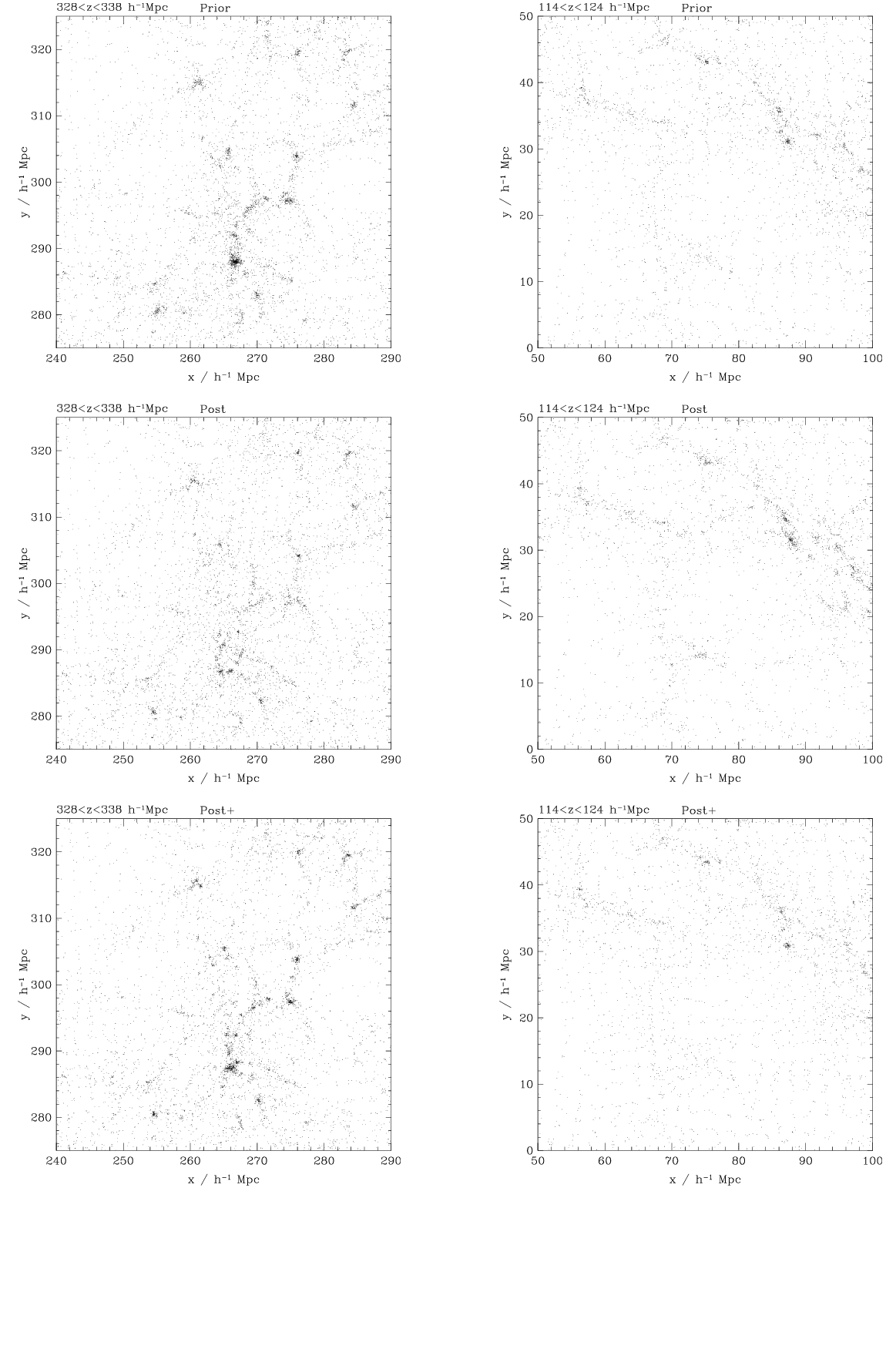}}
\caption{The left hand column of panels show small slices of the Prior, Post 
and Post+ simulations approximately centred on the position where the
long-wavelength density field, $\Dl(\x)$, has its maximum. The panels in the
right-hand column are approximately centred on the position where $\Dl(\x)$, 
has its minimum.}
\label{fig:slices}
\end{figure*}

\section{Comparison with Full Non-linear Computation}\label{sec:nbody}

To test the effect of neglecting the linear to non-linear mode coupling
described in Section~\ref{sec:time} we now compare four sets of particle
distributions derived from two N-body simulations.
The first has no long-wavelength power in its initial conditions, but its
evolved particle distribution is perturbed, a posteriori, using both the 
basic MAP detailed in Section~\ref{sec:space} and the additional modification
to this algorithm described in Section~\ref{sec:time}.
The second simulation has the long-wavelength power included in the initial
conditions prior to being evolved. 
The extent to which the basic and extended algorithms succeed
can then be judged by comparing their particle distributions with that
of the second simulation.

\begin{figure*}
\centering
\centerline{\epsfxsize = 15.0 cm  \epsfbox{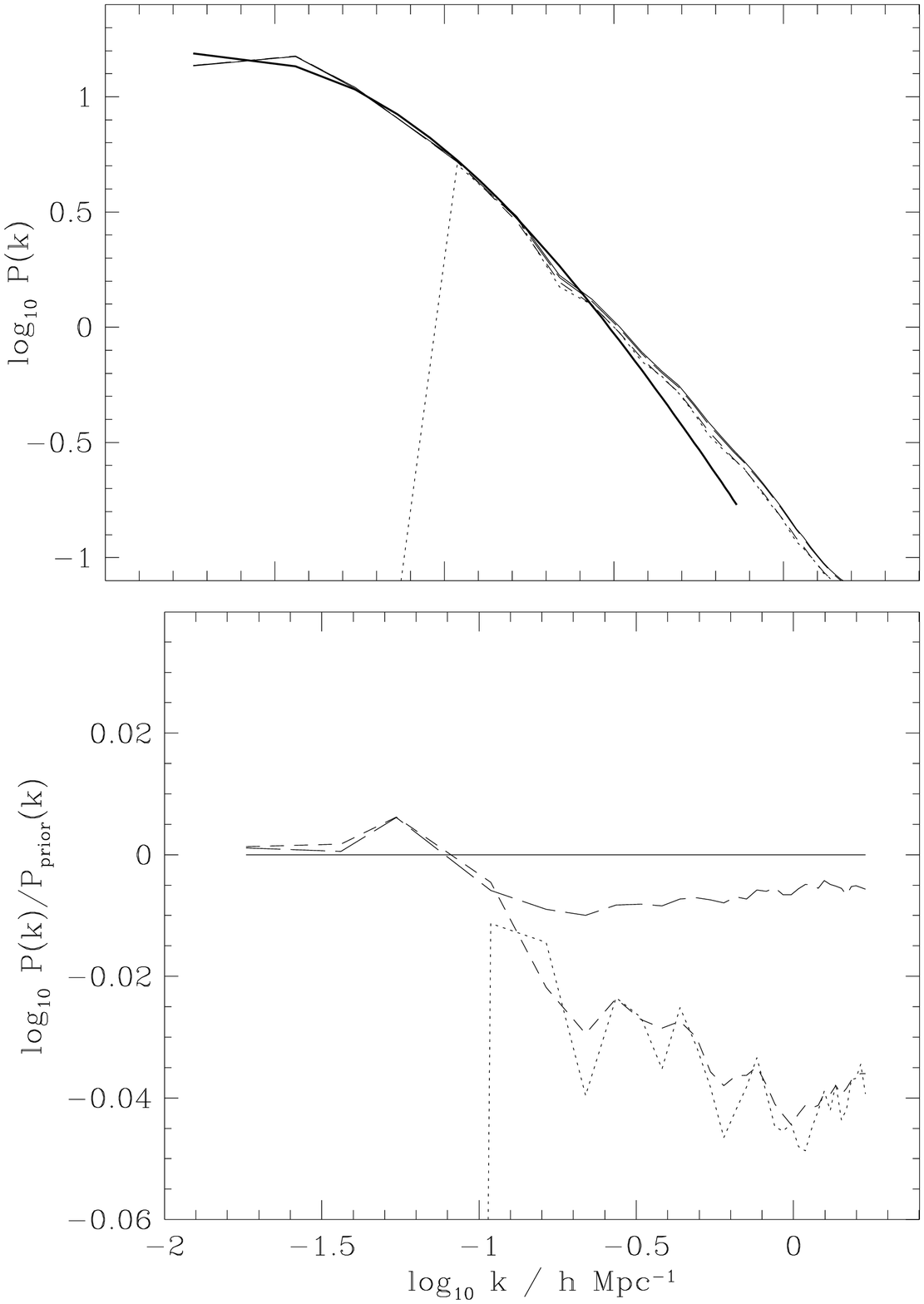}}
\caption{The mass power spectrum, $P(k)$,
in the Prior (solid), Periodic (dotted), Post (short-dashed) and Post+ 
(long-dashed) particle distributions. The upper panel show the $P(k)$,
while the lower panel shows the ratio of the various $P(k)$ to that of the
Prior distribution. Also in the upper panel, 
the heavy solid curve shows the linear theory power spectrum. 
Note, the Post (short-dashed) curve is almost coincident with the 
Prior (solid) curve at small $k$, but
closer to the Periodic (dotted) curve at high $k$. In the
lower panel at high $k$ the  Periodic curve (dotted) can be seen to oscillate
around  Post curve (short-dashed). These small oscillations are probably an
artifact of the way in which the power spectra have been binned.
Each power spectrum has been
averaged in cubical shells of side $2k$ and thickness matched to the
sampling of $\k$-space in the initial conditions of the simulations.}
\label{fig:powermass}
\end{figure*}

\begin{figure*}
\centering
\centerline{\epsfxsize = 15.0 cm  \epsfbox{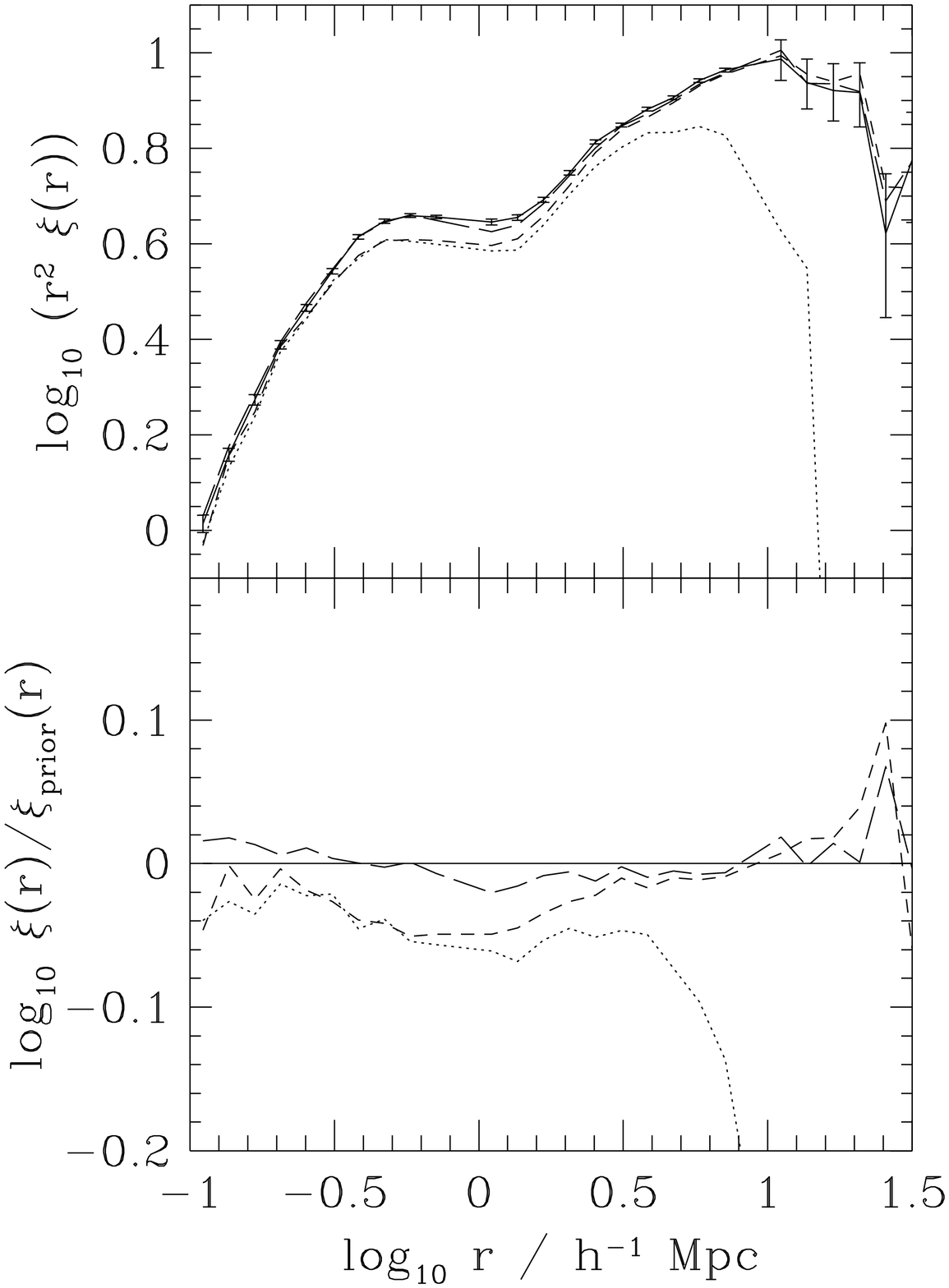}}
\caption{The upper panel show the mass correlation function, 
$r^2 \xi(r)$ estimated from random samples of particles taken from
the Prior (solid), Periodic (dotted), Post (short-dashed) and 
Post+ (long-dashed) particle distributions.
Poisson error bars are indicated on the correlation function of the Prior
simulation. The lower panel shows the ratio of the various $\xi(r)$ to that
of the Prior distribution
so that the differences between the distributions can be seen more
clearly.
}
\label{fig:ximass}
\end{figure*}


\subsection{The Simulations}

We first produced a realization of a Gaussian random field on a
$64^3$ grid with the CDM power spectrum of 
Bardeen \etal \shortcite{bbks} for $\Gamma \equiv \Omega_0 h =0.25$.
We chose the amplitude of the power spectrum such that the linear rms
density fluctuation in spheres of radius  $8 \Mpc$ was
$\sigma_8=0.5$ at the present day, 
in agreement with the value required to produce the
observed abundance of X-ray clusters (\eg White, Efstathiou \& Frenk 1993; 
Eke, Cole \& Frenk 1996; Viana \& Liddle 1996).
The physical size of the box was chosen to be $S = 115.2 \Mpc$ 
and the density parameter was taken to be $\Omega_0=1$.
The only unusual feature about this density field is that we explicitly 
set to zero the power in the cube of 27~modes surrounding $\k=0$.
Thus in the terminology of Section~\ref{sec:space} we choose
$\NS=1$ and explicitly set $\Ds(\x) \equiv 0$ in this density field.

This $64^3$ density field was then used to generate the initial
conditions of two N-body simulations. First the density field was
replicated $3^3$~times to produce a periodic density field on a $192^3$ grid
filling a box of side $L = 3 S =345.6 \Mpc$. This density field and
the Zel'dovich approximation were then used to set up a perturbed grid
of $192^3$ particles. The simulation evolved from these initial
conditions we shall refer to as ``Periodic''. A second set of initial
conditions was generated by adding a density field $\Dl(\x)$ to
the periodic $192^3$ density field.  Here $\Dl(\x)$ was a Gaussian random field
generated on a $192^3$, but with power only in the inner cube of $9^3$
modes. Thus in the terminology of Section~\ref{sec:space}, $\NL = 4$
and the Fourier content of $\Dl(\x)$ covers exactly that region of
$\k$-space left empty in the original $64^3$ density field, but on a
grid which is three times finer.  For the chosen power spectrum and 
normalization the rms value of this field is 
$\langle\vert\Dl(\x)\vert^2\rangle^{1/2} = 0.19$ at the final time.
The simulation run from these
initial conditions in which the long-wavelength density field
$\Dl(\x)$ was added to the particle distribution prior to evolution
we shall refer to as ``Prior''.  
Both simulations were evolved from a starting redshift
of $z=4$ to the present epoch with the AP$^3$M code of Couchman
\shortcite{ap3m}, using a comoving force softening of $\eta = 180
\kpc$ (equivalent to a Plummer softening of $\epsilon=60 \kpc$).

A further particle distribution was generated from the first 
simulation by outputting the position and velocity of each
particle at the time that is satisfied the equation (\ref{eqn:time}).
The long wavelength power, $\Dl(\x)$, was then added to the ``Periodic''
and this latter distribution using the MAP to produce two final particle 
distributions which we refer to as ``Post'' and ``Post+''.


\subsection{The Mass Distribution}

We now compare the mass distributions in the Prior and
Post  distributions. We note that although these simulations do not
exploit the full potential of the MAP (the replicated
volume contains only $64^3$ particles and is replicated only $3^3$ times)
they provide a more direct and stringent test than the simulations
compared in \TB \shortcite{tb}. To the extent to which  
MAP works the Post and Prior particle distributions should be identical.
Thus we are able to study quite subtle differences
between the two simulations that would not have been possible if the
two simulations had different phases and differing resolution as
was the case in \TB \shortcite{tb}.

A visual comparison of the mass distribution in slices through the
two distributions is shown in Figs.~\ref{fig:prior} and ~\ref{fig:post}.
The basic 3-fold periodicity of the small scale structure in the
two distributions is clearly visible. However neither is strictly periodic
due the presence of the long-wavelengths modes, $\Dl(\x)$, present in the 
initial conditions of the Prior simulation
and added by the MAP in the Post distribution.
Clear illustrations how the MAP distorts the underlying periodic
simulation are given in figures~3 and~4 of \TB \shortcite{tb}. 
On first inspection it is impressive how well the Post and
Prior distributions agree. The only discernible differences are on
very small scales.

To study in more detail the differences between the Prior and Post
distributions and also the Post+ distribution we plot, in Fig.~\ref{fig:slices}
expanded views of two different regions. The region plotted in the left-hand 
panels is approximately centred on the location where the long-wavelength 
density field $\Dl(\x)$ has its maximum, while the panels on the right are
at centred on its minimum. For the left hand panels, comparing the Prior
distribution (top) to Post distribution (middle) we see that the Prior
distribution appears more evolved and consists of larger denser clumps.
The converse is true for the Prior and Post distributions shown in the
right-hand panels.
This is in accord with idea that in regions where $\Dl(\x)$ is
positive (negative) structure has evolved more rapidly (slowly)
in the Prior simulation relative to the Periodic or Post simulation.
The extent to which the modification of the MAP algorithm, in which
particles have been output from the N-body simulation at different times,
has succeeded in reproducing this evolution can be judged by studying the
bottom two panels of Fig.~\ref{fig:slices}.
Although not identical, the correspondence between the Post+ and
Prior distributions is very good and much better than that between Post
and Prior.

 The close-up inspection of two of the most extreme regions 
in the simulations has revealed clear differences between the Post
and Prior particle distributions. However these regions are not at all
typical and the impact of these differences on  statistical properties of
the two distributions needs to be assessed.
 We now use the power spectra and correlation functions of the
mass distributions to quantitatively compare the large and small
scale clustering the in the simulations. Fig.~\ref{fig:powermass}
shows the mass power spectra of the four particle distributions.
The power spectra were estimated using an FFT and then averaged in shells.
Normally, one would average the power in spherical shells, but here
we have chosen to average the power in cubical shells so as to match the
geometry of the region of $\k$-space in which the long-wavelength power
has been added. (The oscillations
present in the equivalent figure (fig.~8) of \TB \shortcite{tb}
have been avoided by re-binning the power into bins whose spacing matches
the sampling of $\k$-space in the initial conditions.) 
The first thing to note
is the sharp drop to zero power at $k<0.11 h {\rm Mpc}^{-1}$ 
in the case of the Periodic 
simulation. This corresponds precisely to the region of $\k$-space in which
the power spectrum was set to zero in the initial conditions of the
Periodic simulation. At these same long-wavelengths the power spectra
of the Post and Prior simulation track each other very accurately and
scatter around the required linear theory power spectrum.
At higher-k (smaller scales), all three power spectra lie above the
linear theory power spectrum. Here the power spectrum of the Post simulation
lies closest to that of the Periodic, which is slightly below that of the
Post+ and Prior simulations. These differences can been seen much more clearly
in the lower panel, which plots the ratio of the various power spectra
with respect to that in the Prior simulation.

   We can study the small structure in more detail by examining the
correlation functions of the four distributions shown in Fig.~\ref{fig:ximass}.
The upper panel plots $r^2 \xi(r)$ rather $\xi(r)$ so that an expanded
scale can be used to reveal the quite modest differences between the various
correlation functions. These differences are then high-lighted in the lower
panel which plots the ratio of the various $\xi(r)$ to that of the Prior
distribution.
The lack of large scale power in the Periodic simulation can be seen here
as a dramatic decrease in $\xi(r)$ for $r\gsim 10 \Mpc$. The correlation
function of the Post simulation agrees well with the Prior simulation
on these large scales. At smaller scales, the Post correlation function
peels away from that of the Prior and matches accurately the 
correlation function of the Periodic simulation. The correlation function of
the Post+ distribution lies close to that of the Prior distribution.
This is a convincing demonstration that the basic MAP has essentially worked. 
The small scale structure in the Periodic simulation has been undisturbed 
while at the same time the required large scale power has been generated.
The only sign of the neglect of the interaction between large scale
linear modes and small scale structure is the slight amount by
which $\xi(r)$ for the Periodic and Post simulations lies below that
of the Prior simulation at small scales.


\subsection{The Distribution of Clusters}\label{sec:clusters}

    Here we examine the clustering of a set of objects which do {\it not\ }
simply trace the mass distribution, but are instead biased tracers. 
As an example of such objects we use massive groups or clusters identified
in the simulations. These are strongly biased with respect to the underlying
mass distributions and thus are a demanding test of the algorithms being
studied. For more typical halos, such as galaxy halos, the bias is likely to
be less and differences between the Prior, Post and Post+ distributions
harder to detect and quantify.

\begin{figure*}
\centering
\centerline{\epsfxsize = 10 cm  \epsfbox[40 270 570 774]{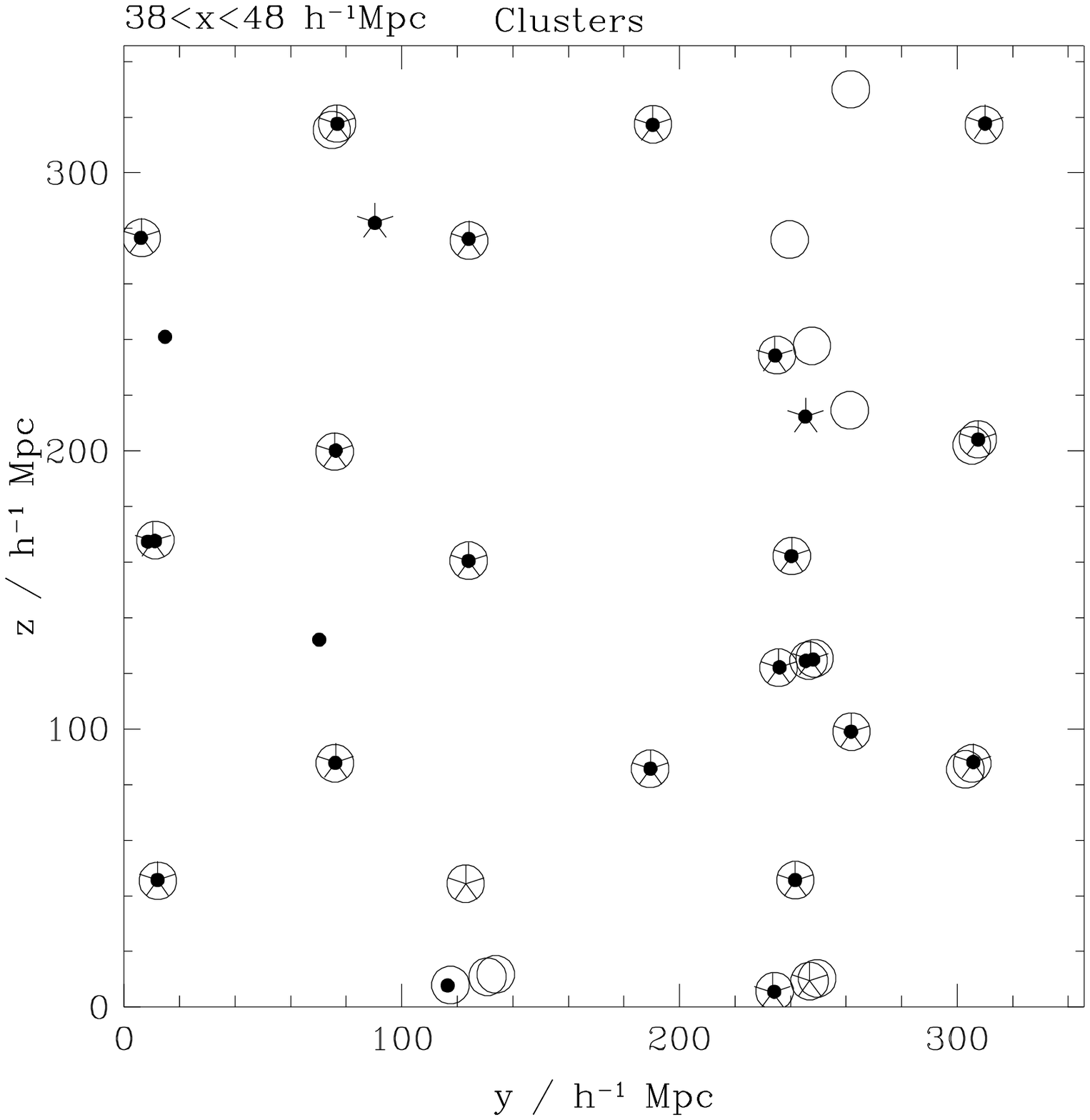} 
\epsfxsize = 10 cm  \epsfbox[40 270 570 774]{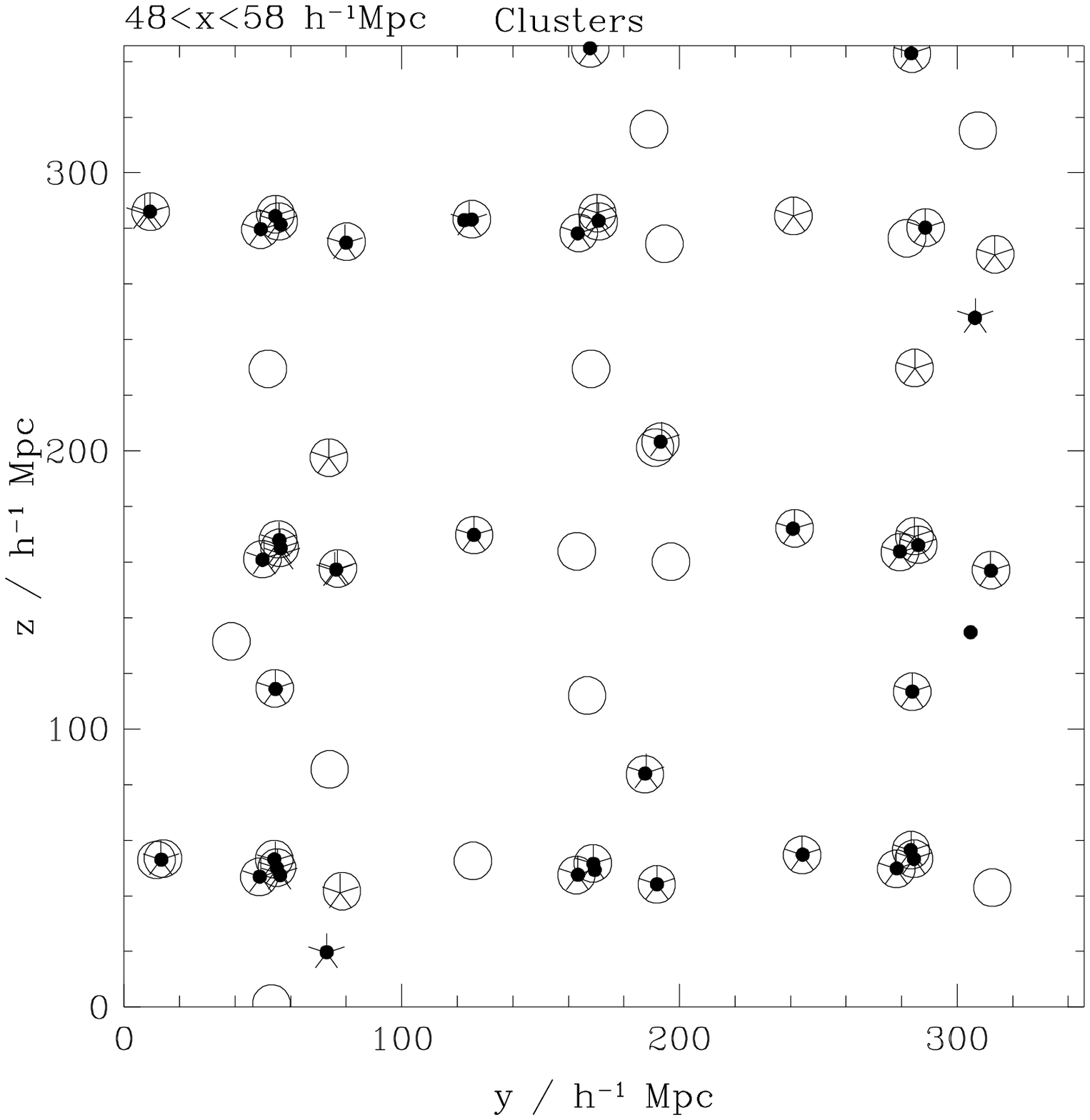} }
\caption{A comparison of the spatial distribution of galaxy clusters as 
defined by the friends-of-friends group finding algorithm in two slices
through the Post (open circles), Prior (filled circles) and Post+ 
(stars) simulations.  
Each slice is $10 \Mpc$  thick and the left-hand slice encloses
the slices of the mass distributions shown in Figs.~\ref{fig:prior} 
and~\ref{fig:post} }
\label{fig:ccdots}
\end{figure*}

    We selected a sample of clusters in each of the simulations using
the standard friends-of-friends group finding algorithm \cite{defw} with a
linking length $0.2$ times the mean inter-particle separation. This
algorithm approximately selects groups with mean over-density of 
$200$.  We then ranked the clusters by mass and retained the $1000$
with highest mass in each simulation. This gives a sample of clusters
with mean separation of $34 \Mpc$ which makes them approximately three
times more abundant than Abell $R \ge 1$ galaxy clusters.  Slices through
these cluster samples are shown for the Prior, Post and Post+ distributions 
in Fig.~\ref{fig:ccdots}. 

In many cases the same cluster is identified in all three
distributions. There are some mismatches where clusters in the Prior
distribution are not found in either the Post or Post+ distributions.
In general the Post+ results match best with the Prior Cluster distribution. 
The most significant difference is the tendency
for the large voids in the Prior and Post+ cluster distribution 
to be peppered
with one or two isolated clusters in the Post distribution. This
can result in a large difference in their large scale clustering 
properties.

\begin{figure}
\centering
\centerline{\epsfxsize = 8 cm  \epsfbox[50 330 550 750]{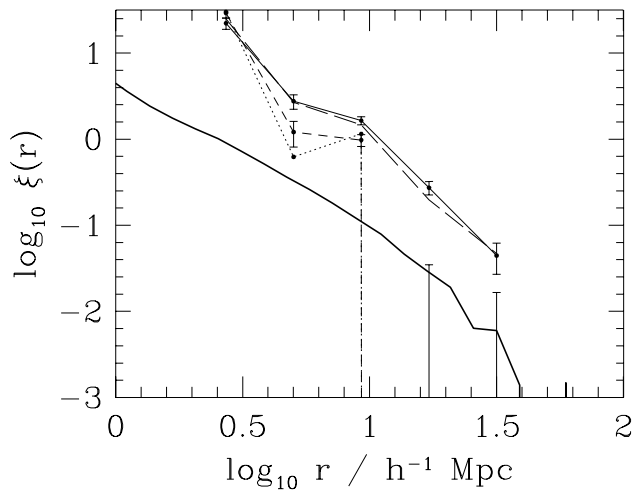}}
\caption{The cluster correlation function, 
$\xi(r)$,  for the Prior (solid),
Periodic (dotted), Post (short-dashed) and Post+ (long-dashed)
distributions. 
Poisson error-bars are indicated on both the
Post and Prior correlation functions. The lower heavy solid line shows
the correlation function of the underlying mass distribution taken from
the Prior simulation.}
\label{fig:xicc}
\end{figure}

\begin{figure}
\centering
\centerline{\epsfxsize = 8 cm   \epsfbox[50 330 550 750]{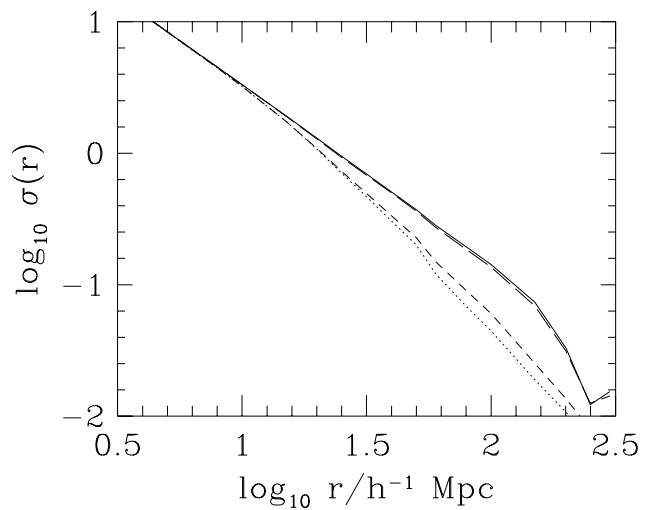}}
\caption{The rms variation, $\sigma(r)$ 
in the number of clusters in spheres of radius $r$ for the Prior (solid)
Periodic (dotted), Post (short-dashed) and  
Post+ (long-dashed) distributions. Note the Prior (solid) and Post+ 
(long-dashed) curves are nearly coincident on at all scales.}
\label{fig:sigma}
\end{figure}

  We examine the clustering properties of the three cluster samples
in Figs.~\ref{fig:xicc} and~\ref{fig:sigma}. The cluster samples are too
small to estimate $\xi(r)$ on very small scales. On large scales we see
that the cluster correlation function in the Prior distribution
is of the same shape as that of the underlying mass distribution
but is offset in amplitude. In contrast the cluster correlation function
in the Post simulation is considerably weaker. The correlation function
for the Post+ distribution agrees well with that of the Prior cluster
distribution on all scales.
The differences in clustering
on large scales are perhaps more clearly seen in Fig~\ref{fig:sigma}, which
plots the fractional rms variation in the number of clusters
in spheres as a function of radius, $r$.

\begin{figure}
\centering  
\centerline{\epsfxsize = 8 cm   \epsfbox[80 420 430 750]{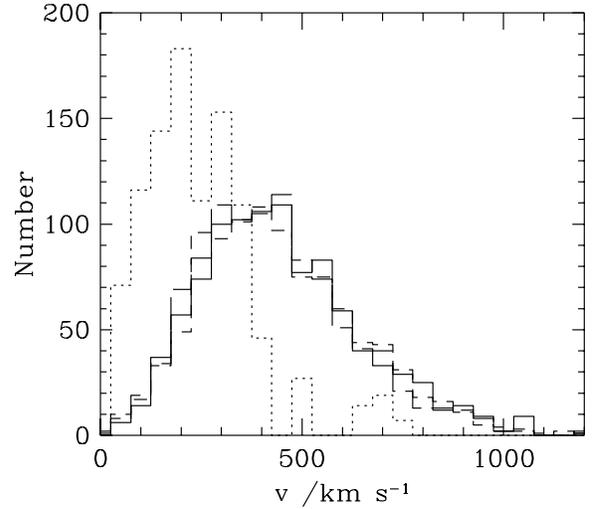}}
\caption{The peculiar velocity distributions in the Prior (solid),
Periodic (dotted), Post (short-dashed),and  Post+ (long-dashed) 
cluster distributions.}
\label{fig:vdist}
\end{figure}

Another comparison of the properties of the cluster
samples is given by the peculiar velocity distribution shown in 
Fig.~\ref{fig:vdist}. In the Prior, Post and Post+ distributions
the large-scale 
velocity fields are very almost equal.
Nonetheless, differences in the peculiar velocity distributions
could arise because of the differing way the cluster populations sample
the velocity field in the two simulations. The similarity of the Post
and Prior distributions shown in Fig.~\ref{fig:vdist} indicate that
this is a small effect in this case, but it is hard to dismiss in general. 
Also note that the peculiar velocity distribution in the Periodic
simulation is much narrower. This demonstrates that long-wavelength modes
produce a large contribution to the peculiar velocity and their effect
is well modelled using the MAP.

The differences we have found in the properties of the cluster distributions
are due to the way long-wavelength modes modulate the cluster abundance 
and in general the properties of the non-linear density field.
In the Prior simulation the cluster abundance is modulated by, and in 
phase with, the long-wavelength modes in the simulation. In contrast
in the Post simulation the cluster abundance is uncorrelated with the
long-wavelength modes which are added in after the simulation is evolved.
This modulation of the cluster abundance is recovered well by the 
additional procedure, described in Section~\ref{sec:time},
incorporated in the Post+ model.


\section{Mock Galaxy Catalogues}\label{sec:mgc}

The differences that we have noted in the cluster catalogues arise 
because the basic MAP does not produce the correct correlation between 
the large-scale
density field and the small-scale non-linear structure.
This results in large differences on large scales for the clusters,
because the cluster selection is done using the small-scale properties of 
the non-linear density field. The same problems will also occur
when constructing mock galaxy catalogues where the galaxy identification or 
biasing algorithm is a function of the small-scale non-linear density field.
Although the magnitude of the problem may be much less if the galaxy population
is not intrinsically strongly biased with respect to the mass distribution.
An example of where the resulting error is important is in the determination
of the density parameter $\Omega_0$ from redshift space distortions of the
clustering pattern \cite{k87}.  At large scales, in the linear regime, 
a distortion is expected whose strength is determined by $\Omega_0^{0.6}/b$
and which is independent of scale. For galaxy catalogues constructed from the
density produced by the basic MAP the bias parameter $b$ will be 
a decreasing function of scale. 
Thus in the mock catalogues one would measure a distortion which
indicates that $\Omega_0^{0.6}/b$ increases with scale rather than converging
to the true value.

  As demonstrated by the Post+ simulation presented above 
these problems with the basic MAP can be largely 
remedied by the additional procedure
described in Section~\ref{sec:time}. However, simpler procedures maybe
valuable for the specific case of creating mock galaxy catalogues.  
One possibility is to use the MAP as normal but
 select galaxies in the {\it initial\ } conditions, 
\ie from the
initial density field including the added long-wavelength contribution
$\Delta(\x) =\Dl(\x)-\Ds(\x)$.
An example of defining galaxies in the initial
conditions, which has been much used, is the peaks biasing scheme
\cite{wedf}.  Here the long-wavelength modes influence the galaxy
selection probability through the way they modulate the number density
of high peaks in the small scale density field \cite{bbks}. 
This effectively by-passes main deficiencies of the MAP.
A second and more flexible approach (Tormen private communication)
is to select galaxies in the non-linear density field before applying
the MAP and explicitly measure their bias, $b$, with respect to
the large scale density field. One then applies the MAP but
boosts the displacements in equations (\ref{eqn:xremove}) and
(\ref{eqn:xadd}) by the bias factor $b$. The perturbations to the
velocities are not altered.
Galaxies catalogues generated using either of these two methods
will have  the correct bias on large scales. 
Thus like the mass distribution the only error in using the MAP 
is then small and confined to small scales.

Clusters identified in the galaxy distribution will suffer fewer problems
than those selected in the mass distribution.  For example 
if either of the two methods proposed above are used to define galaxy
catalogues in the Post and Prior simulations
then the number of galaxies selected in volumes illustrated in the
two top right-hand panels of Fig.~\ref{fig:slices}
should be the same. Statistically, the only way in which the galaxy 
catalogue generated in the Post simulation will differ from that in 
the Prior simulation is in their velocity fields. In the region shown the
Prior simulation has a quieter or cooler velocity field than the
more evolved region in the Prior simulation. Since the region shown in this
figure is only $0.05$\% of the volume of the whole simulation and is
likely to be the most extreme example of this type of difference it
is unlikely to be important for nearly all
applications of mock galaxy catalogues.

\section{Conclusions}\label{sec:disc}

The algorithm devised by \TB \shortcite{tb} allows the dynamic range
of N-body simulations to be extended to very large scales by adding to
them linear power on very large scales.  However, the \TB
\shortcite{tb} Mode Adding Procedure (MAP) neglects coupling between
long-wavelength linear modes and short-wavelength non-linear
modes. This coupling arises through the way long-wavelength density
modes modulate the rate of evolution of small scale structure. Although the
amplitude of this effect is small it is of the same order as the
amplitude of the long-wavelength density modes. Thus for objects such
galaxy clusters  and perhaps galaxies identified in the non-linear density
field the neglect of this coupling leads to a large error in the
amplitude of their large-scale clustering pattern. On the other hand,
for the mass distribution as a whole we find the basic MAP works very
accurately with essentially no error in the mass correlation function
on large scales and only a small error in the amplitude on small
scales.

The main short-comings of the MAP 
can be avoided by two methods. 
The first is to select objects  in the initial conditions where
the initial conditions include the long-wavelength modes that are to be
introduced using the MAP of \TB \shortcite{tb}. The second is
to boost displacements, but not the velocity perturbations,
applied by the MAP to take account of the known bias, $b$
of the selected population of objects.
Thus mock galaxy catalogues generated by either of these methods
will benefit from the ability
of the MAP to better sample modes of the very large scale density field.
Alternatively, if one requires the full mass distribution then the 
\TB procedure can be extended to explicitly address the coupling
of long-wavelength modes to small scale structure by using multiple
outputs from the evolving N-body simulation as described in 
Section~\ref{sec:time}.

\section*{ACKNOWLEDGEMENTS}
SMC would like to thank Carlos Frenk for useful discussions
and Hugh Couchman for providing a copy of his
AP$^3$M N-body code and giving valuable advice.
SMC also gratefully acknowledges the support of a 
PPARC Advanced Fellowship.


\begin{thebibliography}{}
\bibitem[\protect\citename{Bardeen \etal }1986]{bbks}
Bardeen, J.M., Bond, J.R., Kaiser, N., Szalay, A.S., 1986, ApJ, 304,15
\bibitem[\protect\citename{Couchman }1991]{ap3m}
Couchman, H.M.P., 1991, ApJ.Lett, 368,23
\bibitem[\protect\citename{Davis \etal}1985]{defw}
Davis, M., Efstathiou, G., Frenk, C.S., White, S.D.M.,  1985, ApJ, 292,371
\bibitem[\protect\citename{Eke, Cole \& Frenk }1996]{ecf}
Eke, V.R. Cole, S. Frenk, C.S., 1996, MNRAS 282,263
\bibitem[\protect\citename{Gunn }1995]{sdss}
Gunn, J.E. 1995  BAAS, 186,44
\bibitem[\protect\citename{Kaiser }1987]{k87}
Kaiser, N., 1987, MNRAS, 227,1
\bibitem[\protect\citename{Peebles }1993]{peebles}
Peebles, P.J.E., 1993, Principles of Physical Cosmology. 
Princeton University Press, Princeton, NJ.
\bibitem[\protect\citename{Tormen \& Bertschinger }1996]{tb}
Tormen , G., Bertschinger, E., 1996, ApJ, 472,
\bibitem[\protect\citename{Viana \& Liddle }1996]{vl}
Viana, P.T.P., Liddle, A.R.,  1996, MNRAS 281, 323
\bibitem[\protect\citename{White \etal}1987]{wedf}
White, S.D.M., Davis, M., Efstathiou, G., Frenk, C.S.,  1987, Nature, 330,451
\bibitem[\protect\citename{White, Efstathiou \& Frenk }1993]{wef}
White, S.D.M., Efstathiou, G., Frenk, C.S. 1993, MNRAS, 262,1023
\bibitem[\protect\citename{Zel'dovich }1970]{zel}
Zel'dovich Y.B., 1970, A\& A 5,84
\end{thebibliography}
\end{document}